\documentstyle{article}

\raggedright

 {}{\setlength\labelwidth{1.4em}\leftmargin\labelwidth
 \setlength\parsep{0pt}\setlength\itemsep{10pt}
 \setlength{\itemindent}{-\leftmargin}
 \usecounter{enumi}
 
 \sloppy
 \sfcode`\.=1000\relax}

\def\@cite#1#2{{#1\if@tempswa , #2\fi}}
\def\@biblabel#1{}
\def\x{$\pm$}
\def\ha{H$_\alpha$~}
\def\hb{\relax \ifmmode {\rm H}\beta\else H$\beta$\fi}
\def\hi{\relax \ifmmode {\rm H\,{\sc i}}\else H\,{\sc i}\fi}
\def\hii{\relax \ifmmode {\rm H\,{\sc ii}}\else H\,{\sc ii}\fi}
\def\h2{\relax \ifmmode {\rm H}_2\else H$_2$\fi}

\def\fdg{\hbox{$.\!\!^\circ$}}
\def\farcm{\hbox{$.\mkern-4mu^\prime$}}
\def\farcs{\hbox{$.\!\!^{\prime\prime}$}}

\def\degd#1.#2{ #1\fdg#2 }                 
\def\mind#1.#2{ #1\farcm#2 }               
\def\secd#1.#2{ #1\farcs#2 }               
\def\kms{\hbox{$km s^{-1}$}}

\begin{document} \headheight 4cm
\thispagestyle{empty}
\begin{center}

{\LARGE CONSTRAINTS TO THE MASSES OF BROWN DWARF CANDIDATES FROM THE LITHIUM
TEST\footnote{Based on
observations made with the William Herschel Telescope, operated on the island
of La Palma by the Royal Greenwich Observatory in the Spanish Observatorio del
Roque de los Muchachos of the Instituto de Astrof\'\i sica de Canarias, and on
data collected at the European Southern Observatory, La Silla, Chile.}}

\vspace{1.5cm}

{\large
 Eduardo L. Mart\'{\i}n and Rafael Rebolo \\
\smallskip
Instituto de Astrof\'\i sica de Canarias,
E-38200 La Laguna, Tenerife, Spain
\smallskip
e-mail addresses: ege@iac.es and rrl@iac.es
\par\vskip 0.5cm

\medskip Antonio Magazz\`u

\smallskip

Osservatorio Astrofisico di Catania, Citt\`a Universitaria,
I-95125 Catania, Italy

\smallskip
e-mail address: antonio@ct.astro.it}

\bigskip
\bigskip
\bigskip
Send offprint requests to: Eduardo L. Mart\'{\i}n
\end{center}

\newpage

\headheight 45pt \parskip=3mm

\centerline {\large ABSTRACT}

We present intermediate dispersion (0.7-2.2 \AA ~pix$^{-1}$) optical
spectroscopic observations
aimed at applying the ``Lithium Test'' to a sample of
ten brown dwarf candidates located in the general field, two in
young open clusters, and two in close binaries.
We find evidence for strong Li depletion in all
of them, and thus infer lower mass limits of 0.065~M$_\odot$, depending only
slightly ($\pm$0.005~M$_\odot$) on the interior models. None of the
field brown dwarf candidates in our sample appears to be a very young (age
$<$~10$^8$~yr) substellar object. For one of the faintest proper motion
Pleiades members known (V=20.7) the Li test implies a mass  greater than
$\sim$0.08~M$_\odot$, and therefore it is not a brown dwarf. From our spectra
we estimate spectral types for some objects and present measurements
of \ha  emission strengths and radial velocities. Finally, we compare the
positions in the H-R diagram of our sample of brown dwarf
candidates with the theoretical region where Li is expected to be preserved
(Substellar Lithium Region). We find that certain combinations of temperature
calibrations and evolutionary tracks are consistent with the constraints
imposed by the observed Li depletion in brown dwarf candidates, while others
are not.

\par\vskip 1cm
{\em Subject headings:} stars: low mass, brown dwarfs, lithium, spectroscopy --
 binaries: individual (Gl 623, LHS 1047)

\newpage 

\par\vskip 2cm \centerline{\large 1.~INTRODUCTION}

We have proposed that the search for Li in brown dwarf (BD)
candidates can provide a definitive proof of their substellar nature
(Rebolo 1991; Magazz\`u, Mart\'\i n \& Rebolo 1991, 1993; Rebolo, Mart\'\i n \&
Magazz\`u 1992).
Theoretical calculations show that BDs with masses below
0.07~M$_\odot$ preserve a significant fraction of their initial Li content,
while more massive BDs efficiently destroy Li at ages older than about
10$^8$~yr. For higher masses, above the substellar mass limit, total Li
depletion occurs in shorter timescales (Magazz\`u et al. 1993; Nelson,
Rappaport \& Chiang 1993; Bessell
\& Stringfellow 1993; D'Antona \& Mazzitelli 1994).

Many BD candidates have been discovered in the last few years, and
it seems that any dwarf with spectral type later than about M6.5 could
actually have a substellar mass (Kirkpatrick \& McCarthy 1994), depending
on its age.
Isolated field candidates
have uncertain mass estimates based on their position in
the H-R diagram, and on
comparison with theoretical evolutionary tracks. Candidates in binaries
have dynamical constraints on their masses, and can potentially provide a more
reliable proof of a substellar status. However, the uncertainties in
the models and in the binary parameters have so far prevented a definitive
confirmation of a brown dwarf. The search for Li is an independent tool for
probing the masses of BD candidates. Rebolo et al. (1992) coined the term
``Lithium test'' to the use of Li as a discriminant between very low mass
stars and brown dwarfs. The first results of its application on
a small sample of candidates were reported by Magazz\`u et al. (1993).

In this paper we present the results of a Li search in a wide sample
of BD candidates, ranging from proper motion members of the young open clusters
Pleiades and Hyades to some of the coolest and intrinsically faintest field
dwarfs. Preliminary results of this work are presented in Rebolo, Mart\'\i n
\& Magazz\`u (1994).
This paper is organized as follows: Section 2 describes the spectroscopic
observations. Section 3
deals with the analysis of the data, including measurements of \ha
in emission, radial velocities and upper limits to LiI equivalent widths.
Finally, in Section 4 we infer
lower limits to the masses of 14 BD candidates from the observed Li depletions,
and we discuss the implications to current modelling and understanding of the
substellar regime.


\bigskip 

\centerline{\large 2.~OBSERVATIONS}

\begin {table}
\caption{Log of Spectroscopic Observations}
\begin{flushleft}
\begin {tabular}{lccccccc}
& & & & & & & \\
\hline Name  & V & I  & Tel. & $t_{\rm exp}$  & Disp. & Epoch & Ref. \\
   &  &  &  &  (s)  & (\AA\  pix$^{-1}$)  &  & \\
\hline
GL 623 AB     & 10.3 & 8.0 & WHT &  330 & 0.7 & Aug 6, 1992 & 1 \\
LHS 1047 AB   & 11.5  & 8.7    & WHT &  300 & 0.7 & Aug 6, 1992 & 2 \\
HHJ 10 (PPl 10)     & 20.7 & 17.1 & WHT & 8400 &  1.5 & Jan 20, 1993 & 3 \\
LkCa 21     & 13.5 & 11.1 & WHT & 300 & 1.5 & Jan 21, 1993 & 4 \\
V927 Tau    & 14.3 & 11.2 & WHT & 400 & 1.5 & Jan 21, 1993 & 4  \\
BHJ 358        &   &  15.7 & WHT & 3600 & 1.5 & Jan 21, 1993 & 5   \\
LHS 2065     & 18.7 & 14.5 & WHT & 3600 & 1.5 & Jan 21-22, 1993 & 6 \\
LHS 2243     & 19.3: &     & WHT & 3600 & 1.5 & Jan 21, 1993 & 7  \\
LHS 2924     & 19.7 & 15.3 & WHT & 3600 & 1.5 & Jan 21, 1993 & 6  \\
LHS 248 (GJ 1111)     & 14.8 & 10.5 & WHT & 1100 & 1.5 & Jan 21, 1993 & 1 \\
CTI 115638.4+280000 & 20.4 & 16.7 & WHT & 7200 & 1.5 & Jan 22, 1993 & 8 \\
ESO 207-61  & 20.4 &  16.2 & ESO &  7200    &  2.2   &   May 18, 1993 & 6  \\
LHS 2397a  &  19.6 &  14.9 & ESO &  5400    &  2.2   &  May 18,  1993 & 6   \\
Sz 81  &  15.6  &      & ESO &  120    &  2.2   &  May 18, 1993 & 4  \\
TVLM 513-46546  &  &  15.1 & ESO &  4800    &  2.2   &  May 18, 1993 & 9 \\
TVLM 868-110639  & & 15.8 & ESO &  4800    & 2.2    & May 18, 1993 & 9 \\
GL 644 C (VB 8) &  16.8 &  12.2 & ESO &  2600    &  2.2   &  May 18, 1993 & 1
\\
\hline
\end{tabular}
\end{flushleft}
\vskip 2mm
References for V and I: 1. Leggett 1992; 2. Ianna et al. 1988; 3. Stauffer et
al. 1994; 4. Herbig \& Bell 1988; 5. Bryja et al. 1994; 6. Ianna 1993; 7.
Kirkpatrick 1994 (private communication); 8. Kirkpatrick et al. 1994; 9. Tinney
et al. 1993.
\end{table}

Spectroscopic observations have been carried out on one night in August 1992
and during two nights on January 1993
at the 4.2~m William Herschel Telescope (WHT), located at the Observatory del
Roque de los Muchachos on the island of La Palma, and on one night in May 1993
at the 3.6~m telescope of the ESO La Silla Observatory.
The observing log is shown in Table~1, in which the objects are listed by
chronological order of the observations.  The instrumentation
used was the ISIS
double arm spectrograph on the WHT, and EFOSC~1 at ESO. The nominal
dispersions obtained with each instrument are also listed in the Table, and the
slit projection was typically 2 pixels. We obtained the
following spectral coverages: for the stars observed with ISIS on August 1992,
$\lambda\lambda$6370-7270; for the stars observed with ISIS on January 1993,
$\lambda\lambda$5530-6970 (blue arm), $\lambda\lambda$7860-8760 (red arm);
and for the stars observed with EFOSC~1 four orders were recorded
simultaneously, giving $\lambda\lambda$6060-7030, $\lambda\lambda$6780-7910,
$\lambda\lambda$7720-9010, and $\lambda\lambda$8230-10340.

The sample consists of the following objects: HHJ~10, a BD candidate in the
Pleiades cluster, first found in the photometric survey of
Stauffer et al. (1989, their object number 10) and confirmed as a proper motion
member by Hambly, Hawkins \& Jameson (1993);
BHJ~358, the faintest BD candidate
proper motion member of the Hyades in the sample of  Bryja, Humphreys \& Jones
(1994); ESO 207-61, the Hyad moving group candidate discovered by Ruiz,
Takamiya \& Roth (1991);  four of the coolest high proper motion dwarfs known
in the Luyten (1979) LHS catalogue;  CTI 115638.4+280000 (hereafter
CTI~1156+28), a very cool field dwarf from Kirkpatrick et al. (1994); two TVLM
very low-luminosity dwarfs from Tinney, Mould \& Reid (1993);  LHS~248 and
GL~644~C, two benchmarks of cool dwarfs; two binary
systems with BD candidate secondaries (GL623 and LHS1047); and finally
three of the latest M-type T Tauri stars in the Herbig \& Bell (1988)
catalogue.

Each individual spectrum was reduced by a standard procedure using
IRAF\footnote{IRAF is distributed by the National Optical Observatory, which is
operated by the Association of Universities for Research in Astronomy, Inc.,
under contract with the National Science Foundation.}, which included
debias, flat field, optimal extraction and wavelength calibration
using arc lamps. The WHT spectra were finally
flux calibrated using the standard stars Hz~15, Feige~25,
Feige~98 and HD~19445,
which have absolute flux data files available in the IRAF environment.
In Figures~1 and 2 we present the final spectra of ten programme objects, which
are considered to be representative of the whole sample.

\bigskip 

\centerline{\large 3.~ANALYSIS}

The spectral coverage of our data is large enough to
estimate spectral types for those stars in our sample for which there was no
classification in the literature. Only the two close binaries Gl~623 and
LHS~1047,
observed at higher resolution, do not have sufficiently large wavelength
range, and hence they are left out from the analysis of spectral types.
On the other hand, our relatively high dispersion allow us to measure \ha
emission strengths and radial velocities. The spectral types, \ha  equivalent
widths and radial velocities are useful for
interpreting the results of the Li search.

a) Spectral Types

Mart\'\i n (1993) studied different spectral indexes that can be measured
in our spectra, including those of Kirkpatrick, Henry \& McCarthy (1991) and
Prosser, Stauffer \& Kraft (1991). He defined the following index:

$$ M = \frac{F_\lambda(\lambda8155-\lambda8175
)}{F_\lambda(\lambda7875-\lambda7895 )} $$

where the F$_\lambda$s are the average fluxes over these
20~\AA ~wide wavelength intervals.
The M-index provides a smooth second order polynomial correlation with spectral
type in the range M3-M9, the rms being less than half a spectral subclass.
The referee (J.D. Kirkpatrick) has noted that there is a telluric water
band between $\lambda$8164 and $\lambda$8177, and a strong VO bandhead
coming in around $\lambda$7880. We have not attempted to correct our
spectra from telluric lines, but these were not conspicuous in any of the
stars observed. We estimate that the contribution of telluric lines to the
M-index in our spectra was less than 5\% .
We believe that the M-index is mainly sensitive to the slope of the
pseudo-continuum and the strength of VO absorption.
Our spectra in the spectral region where the M-index is defined can be seen
in Mart\'\i n (1993).
For calibrating this index we used programme stars
with spectral types listed
in Herbig \& Bell (1988) and Kirkpatrick et al. (1991, 1994); namely,
LkCa 21 (M3), V927 Tau (M5.5), CTI 1156+28 (M7), GL~644 C (M7), LHS 2397a (M8),
LHS~2243 (M8), LHS~2924 (M9) and LHS~2065 (M9). In Table~2 we present the
spectral types obtained for the rest of our sample.

\medskip

b) \ha  and radial velocities

All our spectra include \ha , which is
seen in emission in most programme stars (see for example Figures~1 and 2).
Our moderately high spectral resolution
has allowed us to detect  weak \ha  emission in several very cool field dwarfs,
including LHS~2924.
The measured equivalent widths of \ha emission
are presented in Table~2. The strength of \ha
could be an age indicator, as it seems to decay from
the Pleiades to the Hyades, and from these to the field dwarfs
for spectral types M0-M5 (Prosser et al. 1991).  However,
note that the \ha emission of BHJ~358 (a Hyad proper motion member) is stronger
than that of the younger Pleiad proper motion member HHJ~10, suggesting that
for spectral types later than M5 \ha may not be
a good indicator of age. It is also worth recalling that \ha can be
highly variable in stars cooler than M5, as observed by us in LHS~2065, and by
other authors in different stars (Bessell 1991, Rebolo et al. 1992). It would
be desirable to monitor extreme M-stars for variability before trying
to use \ha emission as an age indicator.  The possibility that \ha
emission may sometimes be enhanced due to activity in close binaries
should also be checked.

\begin {table} \caption{\ha  and heliocentric radial velocities}
\begin{center} \begin {tabular}{llrl}
\hline Object & Sp.T. & \ha  (\AA~) &  Vrad (\kms)  \\
\hline
LkCa 21       & M3 & 5.1 &  +2 \\
V927 Tau        & M5.5 & 7.2 & +21  \\
Sz 81        & M5.5 & 64.6 & -21:  \\
HHJ 10     & M5.5 & 5.7 & -3 \\
BHJ 358        & M6 & 6.8 & +42 \\
LHS 248           & M7 & 6.0 & -6 \\
CTI 1156+28    & M7 & 2.7 & -19 \\
GL 644 C    & M7 & 5.2 & -13: \\
TVLM 868-110639      & M7.5 & 6.8 &  -32: \\
TVLM 513-46546      & M8 & 2.5 & +31: \\
ESO 207-61      & M8 & 1.9 &  +103: \\
LHS 2243      & M8 & 1.3 & +11 \\
LHS 2397a      & M8 & 22.0 &  -24: \\
LHS 2924          & M9 & 1.8 & -56  \\
LHS 2065          & M9 & 7.5  & +5 \\
                 &    & 20.3  & +4 \\
\hline
\end{tabular} \end{center}
\vskip2truept
Notes: The uncertainties in these quantities are the following: spectral
type \x 0.4,  \ha  equivalent width  \x 0.5 \AA , and
 radial velocity  \x 18 \kms , except for the values marked
with a colon, which have an error bar of \x 30 \kms .
The two values of \ha EW and radial velocity for LHS 2065 correspond to
observations on two different nights.
\end{table}

The  \ha  emission line is the best feature to measure radial velocities in
our spectra as it is the only narrow feature present.  The procedure to derive
the radial velocity was as follows: i) the \ha   emission profile
was fitted by a gaussian function and the center wavelength was measured;
ii) the measured wavelength shifts with respect to the
\ha   laboratory position
were converted into instrumental radial velocities; iii) instrumental
radial velocities were checked so that they were all at the same zero-point
frame by measuring the positions of ten sky lines in each spectrum; iv)
the radial velocities were corrected from diurnal, barycentric and
annual velocities to yield heliocentric radial velocities.
 Despite the fact that \ha emission may form in a region of peculiar
kinematics, we assume
that the radial velocities derived in this way are good enough, given
our rather large uncertainties (18-30 \kms ). As consistency checks we note
that LkCa 21  and V927 Tau
had previously published radial velocities of +6 \kms and +19.6 \kms ,
respectively  (Herbig \& Bell 1988), which are
in good agreement with the velocities obtained by us (Table~2).
 Stauffer et al. (1994) report a radial velocity of +2.8$\pm$4 \kms for HHJ~10,
which is consistent with our value of -3$\pm$18 \kms .
Note that LHS~2065 has two very different
values for the \ha  equivalent width, corresponding
to observations made on January 20 and 21, 1993.
In spite of the large variability of the \ha  strength, the radial velocities
obtained are the same.

The radial velocities of BHJ 358 and HHJ 10 are consistent within the
error bars with those of known members of the Hyades and Pleiades clusters,
which are 30-46 \kms (Kraft 1965) and 0-14 \kms (Stauffer et al. 1984),
respectively. This result reinforces the likelihood of membership to the
clusters.
It is remarkable that the highest
radial velocity in our sample corresponds to ESO~207-61. A high radial
velocity is also found for LHS~2924. These high velocities suggest that
ESO~207-61 and LHS~2924 may be kinematically old.

\medskip

c) Search for LiI

We have inspected the data for the presence of the LiI$\lambda$6707.8 \AA~
resonance line but failed to detect it. We are hence only able to place upper
limits in our target BD candidates.
This can be clearly seen in Figures~1 and 2 where we show the Li region of
several programme objects.
We checked that we could measure the LiI resonance line of the T Tauri stars
LkCa~21, V927~Tau and Sz~81, observed
with the same instrumental setups as the programme BD candidates, and obtain
the same equivalent widths as measured in high resolution spectra
(e.g. Mart\'\i n et al. 1994). The LiI equivalent widths of these T Tauris
range between 750 and 500~m\AA .
The 2$\sigma$ upper limits given in Table~3 were derived considering the
strongest possible feature that could be present in the region around
$\lambda$6708 \AA ,
taking into account the S/N ratio and resolution of each spectrum.
We note that in three cases we cannot rule out the possibility that there
could be weak Li features just at our detection limit (CTI 1156+28, LHS~2397a
and TVLM 513-46546). Higher spectral resolution and better S/N spectra are
necessary to check if there are weak Li lines in these objects.

\begin {table}
\caption{Upper limits to LiI equivalent widths and Li abundances}
\begin{center} \begin {tabular}{lccc}
\hline Object & T$_{\rm eff}$ & LiI (m\AA ~) & log N(Li) \\
\hline
GL 623 B  & 3200 & $<$700  &  $<$2.0  \\
LHS 1047 B  & 3200  & $<$500  &  $<$1.8  \\
HHJ 10      &  3120 & $<$300 &  $<$1.0    \\
BHJ 358     & 3030  &  $<$300 &  $<$-0.2    \\
CTI 1156+28 & 2940  &  $\le$400 &  $\le$0.6  \\
GL 644 C  & 2940 & $<$200  &  $<$-0.3  \\
LHS 248   & 2940 &  $<$170 &  $<$-0.4  \\
TVLM 868-110639    & 2900   & $<$500 & $<$0.5 \\
TVLM 513-46546    & 2880   & $\le$300 & $\le$-0.1 \\
ESO 207-61    & 2880   & $<$300 & $<$-0.1 \\
LHS 2397a    & 2880  &  $\le$700 &  $\le$1.0    \\
LHS 2243    & 2880  &  $<$50 &  $<$-0.4    \\
LHS 2065    & 2630  &  $<$150 &  $<$-1.0   \\
LHS 2924    & 2630  &  $<$250 &  $<$-0.8   \\
\hline
\end{tabular} \end{center}
Note: The upper limits to the equivalent widths listed above have a 90\%
confidence level.
\end{table}

The correct method for deriving the effective temperatures of extreme M-type
stars  remains controversial.
For instance, the temperature of an M6 dwarf ranges
from 3000~K to 2500~K depending on which calibration is adopted.
The effective temperatures listed in Table 3 have been assigned using the
spectral types in Table~2 and the calibration of Kirkpatrick et al. (1993),
whose work
reports the hottest temperatures among all the published T$_{\rm eff}$
conversions. The choice of this warmer temperature scale yields more
conservative estimates for the  derived upper limits to the abundances,
as the LiI line is expected to strengthen by a factor 2-3 from
3000~K to 2500~K (Pavlenko et al. 1994, in preparation).

As argued by Rebolo et al. (1992) and Magazz\`u et al. (1993),
the LiI resonance feature is expected to
be quite strong in dwarf atmospheres with T$_{\rm eff} \sim$2500~K,
and log~N(Li) in the range 3.0-0.0.
 The presence of a strong LiI doublet in the M6 T Tauri star
UX~Tau~C (Magazz\`u et al. 1991) proves
that Li is easily detectable in late M-type stars when its abundance is high.
Furthermore, optical and IR spectra (c.f. Tinney et al. 1993) of the coolest
objects in our sample like LHS~2924, show the presence of many strong atomic
lines from the neutral alkali elements K and Na, whose
atomic structure is similar to Li.
Consequently, we attribute non-detections of the LiI resonance feature in terms
of physical depletion of the photospheric Li abundance.

We did not resolve the secondaries of GL~623 and
LHS~1047, but we estimated the flux of each component
around $\lambda$6707 \AA ~ using the photometric information available, and
corrected the equivalent width upper limit from the contribution of the
primary. Using the new NLTE curve-of-growth calculations and synthetic profiles
presented  in Pavlenko et al. (1994, in preparation), we inferred the upper
limits to the Li
abundances given in Table~3.
Pavlenko et al. show that a BD with
T$_{\rm eff}$=3000~K, which has retained its initial Li content of Log~N(Li)=3,
has EW(LiI)=2.8~\AA . For temperatures cooler by 500~K their results show
that the LiI line is enhanced by about a factor ~3. It is clear that such
strong lines would be very conspicuous, and easily detectable in our spectra.

\bigskip


\centerline{\large 4.~LITHIUM BURNING AND MASS LOWER LIMITS}

The upper limits on the Li abundances of all the BD candidates in our sample
are well below log N(Li)=3, which is the cosmic Li value in the solar
neighbourhood (e.g. Mart\'\i n et al. 1994).
These limits imply large Li depletions
of one or several orders of magnitude. Such an observed depletion constrains
the minimum mass of our programme stars.
Since our sample of BD candidates is heterogeneous, for clarity  we will
discuss our results in three independent subsections, as follows:

\medskip
\centerline{\it 4.1~Isolated Objects in the Field}

An effective way of finding nearby very cool dwarfs  has been
the photographic surveys of high proper motion stars (e.g. Luyten 1979).
We have observed four of the coolest objects known in the LHS catalogue,
namely, LHS~2397a, LHS~2243, LHS~2065 and LHS~2924. They have spectral types
in the range M8-M9 V (Kirkpatrick, Henry \& Liebert 1993), and LHS~2924 has the
lowest luminosity among them (Tinney et al. 1993). Our measurements of a
relatively high radial velocity and weak
\ha  emission suggest that LHS~2924 may be quite old. Its luminosity, lower
than that of LHS2065, may not reflect a meager mass, but an older age
and lower metallicity.

Bessell (1991) suggested that LHS~2397a, LHS~2065 and LHS~2924 may be as young
as a few times  10$^8$ yr  because they have smaller proper motions than
bluer stars. While for LHS~2924 our data do not support youth, for the other
two stars their strong \ha  emission and small radial velocity suggest that
they are young.
However, it is not clear how \ha  could be used as an age indicator as we
briefly discussed in Section 3b.
Since at present the age of field BD candidates cannot be determined
reliably, we will adopt the conservative view that they are old enough to have
burnt Li if their masses allow for it.
The mass lower limit implied by the observed
Li destruction is slightly model-dependent. There are a number of
computations, and we find the following lower limits for an age of
$10^9$ yr: 0.06~M$_\odot$ (Magazz\`u et al. 1993),  0.07~M$_\odot$
(Bessell \& Stringfellow 1993),  0.065~M$_\odot$ (Nelson et al. 1993),
 0.065~M$_\odot$ (D'Antona \& Mazzitelli 1994). If any of these objects
happen to be younger than $10^9$ yr the mass limit would be shifted to
higher masses, but if they were older the mass limit would not change.

The theoretical cooling track of a 0.06~M$_\odot$ BD runs very close and almost
parallel to the faint end of the main sequence (MS) during several
times 10$^7$ yr.
Given the large uncertainties in effective temperature, the position of
isolated field BD candidates in the H-R diagram is a poor indicator of mass and
age, and cannot distinguish between very low mass stars and young BDs.
This is illustrated in Fig.~3 where we show the MS
(isochrone of age 10$^{10}$ yr for masses larger than 0.076~M$_\odot$) and
cooling tracks
of Burrows et al. (1993), corresponding to their model X, and we also mark the
``Substellar Lithium Region''(SLR). We define such a region as that occupied by
objects whose mass is low enough for Li to be preserved with an abundance
log N(Li)$\ge$2.0, i.e. the depletion is less than a factor 10.
In Fig.~3 we have only shadowed the SLR for ages older than 3$\times$10$^6$ yr
because Burrows et al. 1993 do not give younger isochrones. For the purpose of
our study it is relevant to remark that
at ages younger than about $10^8$ yr the SLR includes all the substellar
domain, but for older ages it is restricted to masses below  0.065~M$_\odot$.

The BD candidates
of our sample in common with Tinney et al. (1993) are plotted in Fig.~3,
denoted by crosses, using the luminosities and temperatures given by those
authors.
It is clear that the position of these objects in the SLR
is incompatible with the Li destruction inferred from our observations.
This problem can be solved if we shift the objects towards higher
effective temperatures. We plotted in the Figure (open pentagons) the
same objects but using temperatures estimated from the calibration of
Kirkpatrick et al. (1993) (see Table~3).
In this case the points lie below the MS, and thus
 this calibration gives excessively hot temperatures. The temperature
calibration that would best fit the constraints of Li depletion and theoretical
tracks in the H-R diagram  is intermediate between the one adopted by Tinney et
al. (1993) and Kirkpatrick's et al. (1993).
We have also tested the parameters given by Bessell \& Stringfellow (1993)
for the faintest dwarfs. Their objects in common with our sample are
plotted as black squares in Fig. 3, and we see that the agreement
is better. Only one object (LHS 248) out of seven lies in the forbidden SLR,
and some of them overlap with the predicted position of the MS for masses
 0.09-0.08~M$_\odot$. One of the objects (ESO 207-61)
has a luminosity grossly displaced below the MS, but this may be due
to low metallicity.
We conclude that the temperature calibration adopted
by Bessell \& Stringfellow (1993), which is essentially Bessell's (1991), gives
in general positions in the H-R diagram roughly consistent with the standard
tracks of Burrows et al. (1993), and satisfying the constraint imposed by the
Li test that they cannot be in the SLR since they do not exhibit any lithium.
An important consequence is that
none of the fields BD candidates in Fig.~3 is a contracting very young
substellar object. Nevertheless, it is not discarded that they may be
BDs with masses in the range 0.08-0.065~M$_\odot$ and ages
between 10$^8$~yr and 10$^9$~yr. Objects of such masses and ages are capable of
efficiently destroying Li and yet they will fail to settle on the main
sequence.

\medskip

\centerline{\it 4.2~Objects in Binary Systems}

Very low mass stars in binary systems allow the possibility of obtaining
dynamical information on their masses. The lowest mass eclipsing binary
known to date is CM Dra (GL 630.1), which is composed of two M4.5 dwarfs
of masses 0.237~M$_\odot$ and 0.207~M$_\odot$ (Lacy 1977). Unfortunately
no eclipsing binary system with lower mass dwarfs has ever been discovered.
Other binary systems have dynamical mass measurements of considerably lower
accuracy. Therefore, the application of the Li test sheds light on the
masses of the components.

Magazz\`u et al. (1993) applied the Li test to the binary systems
Wolf 424 (GL 473) and Ross 614 (GL 234), and described how non-detection
of Li constrains the mass range allowed by the orbital error bars.
In this paper we present similar results on GL~623 and LHS~1047. The first one
has two independent estimates of the secondary dynamical mass;
0.067-0.087~M$_\odot$ (Marcy \& Moore 1989), and 0.114$\pm$0.042~M$_\odot$
(Henry \& McCarthy 1993).
The Li depletion inferred by us supports that the mass of GL 623B is
larger than  0.065~M$_\odot$.

The dynamical mass of LHS~1047B is poorly determined; Ianna, Rohde \& McCarthy
(1988) obtained 0.055$\pm$0.032~M$_\odot$, and Henry \& McCarthy (1993) note
that the uncertainties are much larger than $\pm$0.032~M$_\odot$.
The Li non-detection implies that the mass is larger than  0.065 M$_\odot$,
and hence it constraints the range of masses allowed by the error bar.
We note that the results
of the Li test for these binaries and those reported by Magazz\`u et al. (1993)
are consistent with the finding of Kirkpatrick \& McCarthy (1994) that
the secondaries of Ross~614 and LHS~1047 have spectral types no later
than M6.5 V.

\medskip
\centerline{\it 4.3~Isolated Objects in Open Clusters}

Several searches for substellar objects in nearby open clusters have been
carried out in the last few years (e.g. Stauffer et al. 1989, Bryja et al.
1994). Dozens of BD candidates were identified
on the basis of very red optical or IR colours, but for only very few
of them do proper motion and spectroscopic studies exist. We have
chosen to apply the Li test to one of the reddest proper motion members
known in the Pleiades (HHJ~10, Hambly et al. 1993), and the faintest
proper motion Hyades member in the sample of Bryja et al. (1994).
Our analysis of spectral types, \ha emission and radial velocity presented
in Section~3 confirm that HHJ~10 and BHJ~358 are {\it bona-fide} members
of these clusters.

The Li test provides stronger mass constraints in Pleiades stars than in
Hyades stars because of the difference in ages of about a factor ten.
The SLR has an upper mass limit of 0.08~M$_\odot$ or higher for ages
lower than about 10$^8$ yr, while for ages of a few $\times 10^8$ yr the
upper mass limit is $\sim$0.065~M$_\odot$.
In Figure~4 we compare the H-R diagram position of HHJ~10 with the
theoretical models of Burrows et al. (1993). We have used the photometric data
in Steele, Jameson \& Hambly (1993) and the conversions from I-K colour to BC
and T$_{\rm eff}$ of Bessell (1991). It is clear that HHJ~10 is inside the SLR,
and the tracks indicate a mass of about 0.055~M$_\odot$.  This is obviously
inconsistent with the large Li depletion observed.

Very recently Marcy, Basri \& Graham (1994) applied the Li test to two other
Pleiad BD candidates; HHJ~3 and HHJ~14. The first is about half a magnitude
fainter in I than HHJ~10, and the second  is slightly brighter, but
all three have the same colour (I-K)=3.3 (Steele et al. 1993).
Just one spectral type is available, which is M5.6 for HHJ~10 as derived by us
using the scale of Kirkpatrick et al. (1991).
Marcy et al. (1994) report Li non-detections in their spectra, which have
considerably higher resolution than but similar S/N ratio as ours.

As we pointed out before, the Li test does not confirm the substellar mass
implied by the position in the H-R diagram of HHJ~10.
In Figure~4 we plot HHJ~3 and HHJ~14 in the H-R
diagram and we see that they lie in the SLR as well. In order
to bring the positions of these stars outside the SLR, they would have to
be shifted by 200~K to hotter T$_{\rm eff}$. Current uncertainties
in assigning temperatures to these objects could allow for such an effect.
In particular Steele et al. (1993) have noted that the (R-I) vs. (I-K)
relationship for very low mass Pleiades stars seems to be different than that
of field dwarfs.
We note that our spectral type for HHJ~10 of about M5.5 is more in agreement
with its (R-I) colour of 1.9 (Steele et al. 1993) than with its (I-K) of 3.3 .
Use of (R-I) for deriving a  T$_{\rm eff}$ gives
a value of 3000~K, which is a little more than 200~K hotter than the
T$_{\rm eff}$ derived from (I-K).
Because the spectral type and
(R-I) colour imply a T$_{\rm eff}$ hot enough  to explain the Li burning, this
suggests
that the (I-K) colour for very low mass Pleiades stars may be systematically
redder than for field stars. Such an effect may be caused by a near-IR
excess in these young objects, or anomalous extinction, or
perhaps a peculiar molecular band pattern. Spectroscopic
observations in the optical and the IR are encouraged to establish the
cause of this colour anomaly.

On the other hand it is worth recalling that the theoretical PMS tracks are
quite uncertain. For instance,
D'Antona \& Mazzitelli (1994) have discussed to some extent the influence
of different opacities and treatments of convection.
In particular, we note that their new tracks
with the Canuto-Mazzitelli convection treatment, are hotter by about 180~K at
0.08~M$_\odot$,
and age 7$\times$10$^7$ yr, than model X of Burrows et al. (1993).
Therefore, at the age of the Pleiades the SLR
based on D'Antona \& Mazzitelli's models is shifted towards hotter temperatures
and hence  it is more difficult to explain the observed Li depletion in our BD
candidates.

The discrepancy between the
Li test and the masses implied by the PMS models may be a consequence of
assigning too low T$_{\rm eff}$ to very low mass Pleiades stars, and/or
that the tracks predict temperatures that are too hot for young substellar
objects.
{}From the considerations given above about the (I-K) colours of Pleiades BD
candidates, we favour the hypothesis that the temperatures of
HHJ 3, 10 and 14 are higher by about 200~K than previously thought.
It is necessary to derive reliable  T$_{\rm eff}$ for the lowest
mass Pleiades stars in order to test theoretical evolutionary models.

With respect to the Hyad BD candidates, we consider the cases of BHJ 358
and ESO 207-61. The latter object is possibly a member of the Hyades moving
group according to Ruiz et al. (1991). However, our radial velocity
is very high and the luminosity given by
Bessell \& Stringfellow (1993) is very low, and therefore its connection
to the Hyades seems unlikely. Thus,
we have plotted only the position of BHJ~358 in Figure~4 using
the photometry given by Bryja et al. (1994) and the calibrations of Bessell
(1991). It can be seen that BHJ 358 lies just outside the SLR.
We can only impose a lower limit of 0.065~M$_\odot$ from the observed
Li depletion which is consistent with  the theoretical mass of
$\sim$0.08~M$_\odot$
inferred from its position on the H-R diagram.

\bigskip 

\centerline{\large 5.~CONCLUSIONS}

The central conclusion of this paper is that Li has not been detected in
intermediate resolution spectra of 14 BD candidates. We estimate minimum
Li depletions in these objects ranging from a factor of 10 for GL~623~B to
a factor of 10$^4$ for LHS~2065 and LHS~2924. The fact that Li has been
efficiently burnt implies that the masses are greater than
0.065$\pm$0.005~M$_\odot$.

Our data support membership of HHJ~10 and BHJ~358 to the Pleiades
and Hyades clusters, respectively, because
of their late spectral type, \ha emission and consistent radial velocities. The
non Li detections imply masses $>$ 0.08~M$_\odot$ for the Pleiades
member and $>$ 0.065~M$_\odot$ for the older Hyades member.

In order to illustrate the implications of the Li test in the H-R diagram we
have defined the Substellar Lithium Region (SLR).
The BD candidates
are located inside or outside the SLR depending on the temperatures adopted
for them. Since the observed Li depletions prevent these objects from
being inside the SLR we conclude that some temperature calibrations
give excessively  low values of T$_{\rm eff}$. Only a narrow region in the H-R
diagram, between the SLR and the main sequence, is allowed for our
sample of very low mass dwarfs, and the temperature calibrations should satisfy
these constraints.

At ages younger than about 10$^8$ yr the SLR comprises all the substellar
domain. Thus, young objects, like the Pleiades members, that have
burnt their Li are not brown dwarfs. Adopting the frequently
used (I-K) vs. T$_{\rm eff}$ calibration for the faintest proper motion
Pleiades members currently known, leads to the inconsistent result that
they lie well inside the SLR in the H-R diagram. We argue that
this inconsistency could be solved if these stars are about 200~K hotter
than inferred from their (I-K) colours. The (R-I) colour and the spectral
type of HHJ~10 suggest that such a hotter temperature may be correct.
Further considerations on the validity of theoretical
evolutionary tracks and isochrones have to be postponed
until a reliable temperature scale is established for the very low
mass Pleiades objects.

\medskip

\bigskip
\centerline{\large ACKNOWLEDGEMENTS}

Drs. F. Allard, G. Basri, C. Bryja, F. D'Antona, N.C. Hambly, T.J. Henry, R.F.
Jameson, J.D. Kirkpatrick, G.W. Marcy, C.F. Prosser,  J. Stauffer and
C.G. Tinney are
gratefully acknowledged for communicating results prior to publication.
We thank J.A. de Diego for
his kind help with the English, and the referee (J.D Kirkpatrick) for
many comments that helped us to improve this paper.
This work has been partially supported by the Spanish DGICYT project No.
PB92-0434-C02.

\newpage

\section*{References} \begin{trivlist}

\item [] Bessell, M.S. 1991, AJ, 101, 662

\item [] Bessell, M.S. \& Stringfellow, G.S. 1993, ARAA, 31, 433

\item [] Bryja, C., Humphreys, R.M. \& Jones, T.J. 1994, AJ, 107, 246

\item [] Burrows, A., Hubbard, W.B., Saumon, D.  \&  Lunine, J.I. 1993, ApJ,
406, 158

\item [] D'Antona, F. \& Mazzitelli, I. 1994, ApJS, 90, 467


\item [] Hambly, N.C., Hawkins, M.R.S.  \&  Jameson, R.F. 1993, A\&AS, 100, 607

\item [] Henry, T.J. \&  McCarthy, Jr., D.W. 1993, A.J, 106, 773

\item [] Herbig, G.H. \&  Bell, K.R. 1988, Lick Obs. Bull., No. 1111

\item [] Ianna, P.A. 1993, in ``Developments in Astrometry and their Impact
on Astrophysics and Geodynamics'', p. 75

\item [] Ianna, P.A., Rohde, J.R.  \&  McCarthy, D.W. 1988, AJ, 95, 1226

\item [] Kirkpatrick, J.D. \& McCarthy, Jr, D.W. 1994, AJ, 107, 333

\item [] Kirkpatrick, J.D, Henry, T.J. \& McCarthy, Jr., D.W. 1991, ApJS, 77,
417


\item [] Kirkpatrick, J.D, Kelly, D.M., Rieke, G.H., Liebert, J., Allard, F.
\& Wehrse, R. 1993, ApJ, 402, 643

\item [] Kirkpatrick, J.D, McGraw, J.T., Hess, T.R., Liebert, J., \& McCarthy,
Jr, D.W. 1994, ApJS, in press

\item [] Kraft, R.P. 1965, ApJ, 142, 681

\item [] Lacy, C. 1977, ApJ, 218, 444

\item [] Leggett, S.K. 1992, ApJS, 82, 351

\item [] Luyten, W.J. 1979, LHS Catalogue, Minneapolis, Univ. of Minnesota

\item [] Magazz\`u, A., Mart\'\i n, E.L.  \&  Rebolo, R. 1991, A\&A, 249, 149

\item [] Magazz\`u, A., Mart\'\i n, E.L.  \&  Rebolo, R. 1993, ApJ, 404, L17

\item [] Marcy, G.W. \&  Moore, D. 1989, ApJ, 341, 961

\item [] Marcy, G.W., Basri, G., \& Graham, J.R. 1994, ApJ, in press

\item [] Mart\'\i n, E.L. 1993, PhD Thesis, Universidad de La Laguna

\item [] Mart\'\i n, E.L., Rebolo, R., Magazz\`u, A. \& Pavlenko, Ya.V. 1994,
A\&A, 282, 503

\item [] Nelson, L.A., Rappaport, S.  \&  Chiang, E. 1993, ApJ, 413, 364


\item [] Prosser, C.F., Stauffer, J. \& Kraft, R.P. 1991, AJ, 101, 1361

\item [] Rebolo, R. 1991, in ``Evolution of Stars: The Photospheric Abundance
Connection'', IAU Symp. 145, eds G. Michaud \& A. Tutukov, Kluwer, p. 85

\item [] Rebolo, R., Mart\'\i n, E.L.  \& Magazz\`u, A.  1992, ApJ, 389, L83

\item [] Rebolo, R., Mart\'\i n, E.L.  \& Magazz\`u, A.  1994, in
8th Cambridge workshop on Cool Stars, Stellar Systems and the Sun, ed J.-P.
Caillault, PASP Conf. Series.

\item [] Ruiz, M.T., Takamiya, M.Y.  \& Roth, M.  1991, ApJ, 367, L59


\item [] Stauffer, J., Hartmann, L.W., Soderblom, D.R. \& Burham, J.N. 1984,
ApJ, 280, 202

\item [] Stauffer, J., Hamilton, D., Probst, R., Rieke, G. \& Mateo, M. 1989,
ApJ, 344, L21

\item [] Stauffer, J., Liebert, J., Giampapa, M., Macintosh, B., Reid, N.
\& Hamilton, D., 1994, AJ, in press

\item [] Steele, I.A., Jameson, R.F. \& Hambly, N.C. 1993, MNRAS, 263, 647

\item [] Tinney, C.G., Mould, J.R. \& Reid, I.N. 1993, AJ, 105, 1045

\end{trivlist}

\newpage

\centerline{\large Figure Captions:}

{\bf Figure 1.} Final spectra of five programme stars from Table~1.
The left panel displays a wide
spectral range, and the right panel shows a zoom of the region
around the LiI $\lambda$6707.8\AA ~ resonance line. Fluxes are relative,
with the pseudo-continuum at $\lambda$6700\AA ~ taking a value of unity.
 The resolutions are between FWHM= 1.4 and 4.4\AA ~(see Table~1).
A box-car smoothing of width 2 pixels has been applied. The top spectrum
is that of a T Tauri star, showing a strong LiI resonance line of
EW=715~m\AA ~ (Mart\'\i n et al. 1994). In contrast, the other spectra do not
present detectable LiI lines.

{\bf Figure 2.} Final spectra of another five programme stars from Table~1.
This Figure is organized in the same way as the previous one.
Note the presence of a weak feature in CTI~1156+28 near the expected position
for the LiI line.

{\bf Figure 3.} Our sample of field BD candidates in the H-R diagram.
Crosses denote stars in common with Tinney et al. (1993), plotted using the
luminosities and temperatures reported by them. Open pentagons are the same
objects and luminosities but the temperatures are obtained from the calibration
of Kirkpatrick et al. (1993). Black squares are our stars in
common with Bessell \& Stringfellow (1993), using the parameters given by
them. For comparison with theoretical computations we have superimposed
the tracks of Burrows et al. (1993) for masses of 0.1, 0.08 and
0.06~M$_\odot$ (dashed lines), and we have also drawn their isochrones for 3
$\times 10^6$ yr and 10$^{10}$~yr (solid lines).
The Substellar Li region defined in the text is marked with slanting lines.

{\bf Figure 4.} Faint proper motion cluster members in the H-R diagram. We plot
the Pleiades stars HHJ~3, 10 and 14 as asterisks, and the Hyades star BHJ~358
as an open polygon. The tracks and
SLR are the same as in the previous Figure. We have superimposed the
isochones from Burrows et al. 1993 for 70 Myr (Pleiades age) and 600
Myr (Hyades age). Note that all three Pleiades
BD candidates fall inside the SLR and above the 70 Myr isochrone.

\end{document}